\documentclass[a4paper,11pt]{article}
\usepackage{pos}
\usepackage{subcaption}
\usepackage{natbib}

\title{Associated $W$ + charm production: indications for PDF analysis}

\author*[a]{V. Alanko}
\author[a]{I. Helenius}
\author[a]{H. Paukkunen}

\affiliation[a]{Department of Physics, University of Jyväskylä,
P.O. Box 35, FI-40014 University of Jyväskylä, Finland\\
Helsinki Institute of Physics, 
P.O. Box 64, FI-00014 University of Helsinki, Finland}

\emailAdd{alankovh@jyu.fi}
\emailAdd{hannu.t.paukkunen@jyu.fi}
\emailAdd{ilkka.m.helenius@jyu.fi}

\abstract{The strange quark and antiquark contents of the proton remain weakly constrained compared to the other light quarks, particularly the asymmetry between the two. Production of a $W^\mp$ boson in association with a charmed meson $D^{(*)\pm}$ could provide additional constraints in future fits of parton distribution functions. We calculate this process in general-mass variable-flavor-number scheme at next-to-leading order in perturbative Quantum Chromodynamics. We investigate the production ratio between events with oppositely charged $W$ bosons, in which various theoretical uncertainties largely cancel while sensitivity to PDFs remains. We compare our predictions for this quantity with the recent ATLAS data at $\sqrt{s} = 13$ TeV and find that CT18ANLO, which sets the strangeness asymmetry to zero, yields good agreement with the data while MSHT20NLO and NNPDF4.0NLO, which have positive strangeness asymmetry in the relevant range of momentum fraction $x$, show more tension with the data. An approximate PDF-level analysis of the production ratio indicates that this tension could be attributed to a too large strangeness asymmetry.}

\FullConference{The 33rd International Workshop on Deep Inelastic Scattering and Related Subjects (DIS2026)\\
4 - 8 May 2026\\
Bologna, Italy\\}

\begin{document}
\maketitle

\section{Introduction}
In global determinations of parton distribution functions (PDFs), the strange-quark distribution is less well constrained than those of the other light quarks. In particular, there is no conclusive consensus concerning the size of the the strangeness asymmetry \cite{Anderson:2024evk}, defined as the difference between the strange and the antistrange distributions. The production of a $W$ boson in association with a charm quark, measured by the Large Hadron Collider (LHC) with $\sqrt{s}$ ranging from $7$ to $13$ TeV \cite{CMS:2013wql, CMS:2018dxg, LHCb:2015bwt, ATLAS:2014jkm, ATLAS:2023ibp}, could provide more understanding in this respect.

As the quark in $W^\mp c/\Bar{c}$ production cannot be detected directly, such processes have to be tagged either by a charmed jet or a charmed hadron. The former is known up to next-to-next-to-leading order (NNLO) in perturbative QCD \cite{Czakon:2022khx} while the latter is known at this order only within the zero-mass approximation \cite{Generet:2025bqx}. In our work \cite{Alanko:2026hoc}, we consider $W^\mp D^{(*)\pm}$ production in general-mass variable-flavor-number scheme (GM-VFNS) at next-to-leading order (NLO). We advocate the charge ratio 
\begin{equation}
    \label{eq: Rcpm}
      R_c^{\pm} \equiv \big[{\sigma(W^++D^{-}) + \sigma(W^++D^{*-})}\big]\big/\big[{\sigma(W^-+D^{+}) + \sigma(W^-+D^{*+})}\big],
\end{equation}
in which uncertainties due to missing NNLO corrections and D-meson fragmentation functions (FFs) largely cancel. The sensitivity to charge-asymmetry in the PDFs, however, remains and confronting the calculation with experimental data can therefore give new insights to PDFs. Here, we will compare our results with the recent ATLAS data at $\sqrt{s} = 13$ TeV \cite{ATLAS:2023ibp}.

\section{Production of $W^\mp D^{(*)\pm}$ in GM-VFNS}
To illustrate how our calculation of $W^\mp D^{(*)\pm}$ works, let us consider the case of $W^-D^+$ as an example. The differential cross sections can schematically be written as
\begin{align}
    \label{eq: calculation}
    &\frac{d\sigma(W^-D^+)}{dp_{T,D} d\eta_D d\eta_\text{l}} = \sum_{i, j = s, d, g} f_i(\mu) \otimes \frac{d\hat{\sigma}_\text{NLO}(ij\rightarrow W^- c, \mu)}{dp_{T,c} d\eta_c d\eta_\text{l}} \otimes f_j(\mu) \otimes D_{c \rightarrow D^+} (\mu)\notag\\
    & - \frac{d\sigma_\text{LO}(W^- c, \mu)}{dp_{T,c} d\eta_c d\eta_\text{l}} \otimes \frac{\alpha_s(\mu)}{2\pi} C_F\left[\frac{1+l^2}{1-l}\left(\log\frac{\mu^2}{m_c^2}-1-2\log(1-l)\right)\right]_+ \hspace{-0.2cm}\otimes \hspace{-0.0cm} D_{c\rightarrow D^+}(\mu).
\end{align}
where $p_{T,D}$ and $d\eta_D$ refer to the transverse momentum and pseudorapidity of the D meson, and $p_{T,c}$ and $d\eta_c$ are the corresponding charm-quark variables. The W is considered to decay leptonically and $\eta_\text{l}$ is the pseudorapidity of the charged lepton originating from this decay. The renormalization/factorization scales are collectively denoted by $\mu$. On the first line, the partonic cross sections for $W^- c$ production with initial-state partons $i$ and $j$ are convoluted with PDFs of those partons, $f_i$ and $f_j$. We perform this part of the calculation by utilizing the publicly available MCFM program \cite{Campbell:1999ah, Campbell:2011bn, Campbell:2019dru}. The output is a finely binned $W^- c$ cross section, which we then convolute with the FF $D_{c\rightarrow D^+}$. In GM-VFNS, the FF evolves according to the DGLAP equations \cite{Altarelli:1977zs} which resums collinear radiation of partons. Consequently, the first line in Eq.~\eqref{eq: calculation} contains double counting of contributions where a gluon is radiated collinearly from the final-state charm. The logarithmic term in the second line ($C_F=4/3$) containing the charm-quark mass $m_c$ subtracts this double counting, with the LO $W^-c$ cross section again obtained from MCFM.

The ATLAS collaboration \cite{ATLAS:2023ibp} measured the difference
\begin{equation*}
    \sigma_\text{OS-SS} (W^\mp D^{(*)\pm}) = \sigma (W^\mp D^{(*)\pm}) - \sigma(W^\mp D^{(*)\mp}).
\end{equation*}
Here, the same-sign (SS) events are subtracted from the opposite-sign (OS) events. This is called the OS-SS subtraction. The subtraction largely suppresses the background with no strange in the initial state. MCFM already calculates this specific version of the partonic $W^\mp c/\Bar{c}$ cross section, letting us compare our values directly with the measurement.

\section{Results}
\begin{figure}[h!]
    \centering
    \begin{subfigure}{0.49\linewidth}
        \centering
        \includegraphics[width=\linewidth]{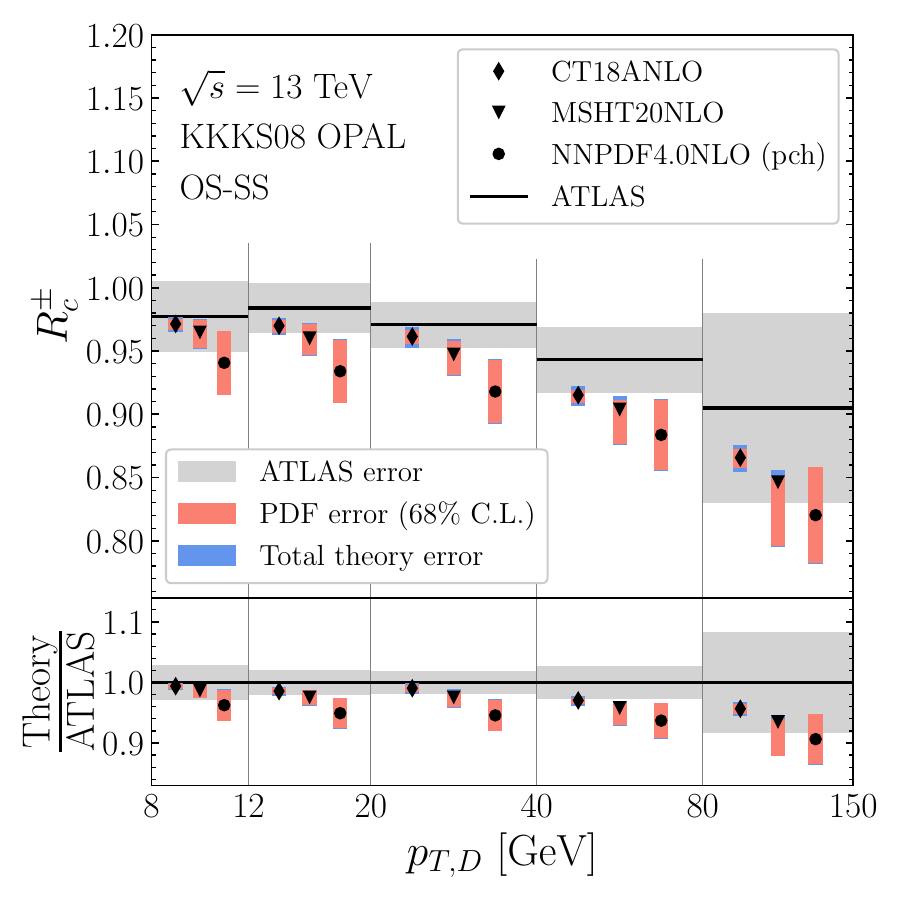}
    \end{subfigure}
    \begin{subfigure}{0.49\linewidth}
        \centering
        \includegraphics[width=\linewidth]{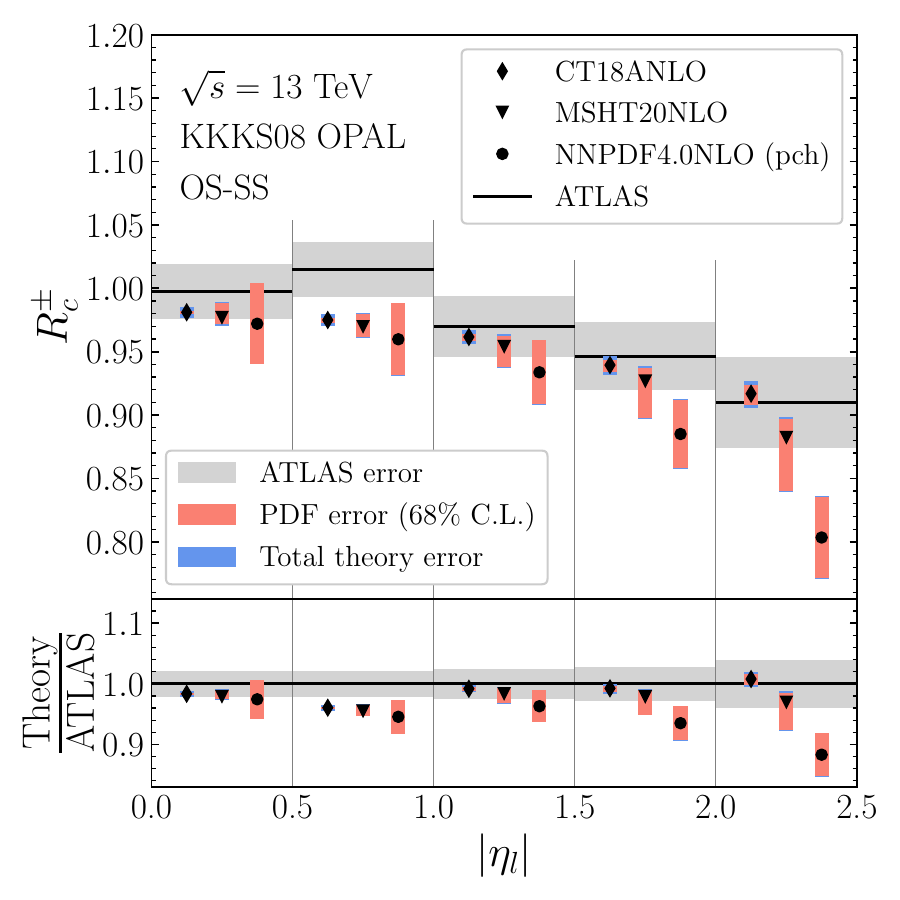}
    \end{subfigure}
    \caption{The production ratio $R_c^\pm$ as a function of 
    $p_{T, D}$ and $|\eta_l|$.
    The calculation uses the KKKS08 parametrization of FFs \cite{KKKS08}.
    }
    \label{fig: Rcpm}
\end{figure}

Figure~\ref{fig: Rcpm} shows the values of $R_c^\pm$ as a function of $p_{T, D}$ and $\eta_l$. The ATLAS data \cite{ATLAS:2023ibp} are shown as the horizontal lines with gray error bands. The theory predictions are provided with three different PDF fits: CT18ANLO \cite{Hou:2019efy}, MSHT20NLO \cite{Bailey:2020ooq} and NNPDF4.0NLO with perturbative charm (pch) \cite{NNPDF:2021njg}. We notice that the scale uncertainty on the theory values, estimated by varying all three scales uniformly up and down by a factor of two, is almost non-existent. Out of the three PDF fits CT18A, which sets $s_\text{valence}$, i.e. the strangeness asymmetry, to zero shows the best agreement with the data. The other two fits have $s_\text{valence} > 0$ in the relevant range of momentum fraction $x \sim 10^{-2}$ and tend to show larger deviations from the data. In general, the theory values fall below the data, and the differences between the three PDF fits grow as $p_{T, D}$ and $|\eta_l|$ grow.

\begin{figure}[t!]
    \centering
    \begin{subfigure}{0.49\linewidth}
        \centering
        \includegraphics[width=\linewidth]{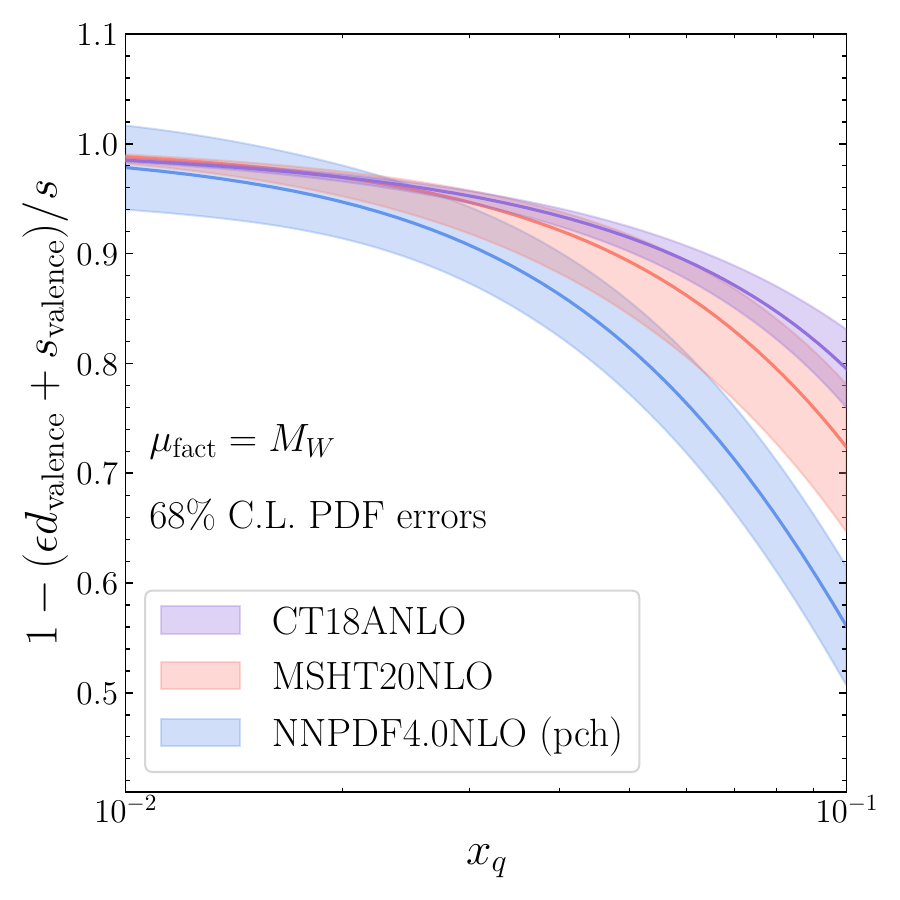}
        \caption{}
        \label{fig: Rcpm approximation}
    \end{subfigure}
    \begin{subfigure}{0.49\linewidth}
        \centering
        \includegraphics[width=\linewidth]{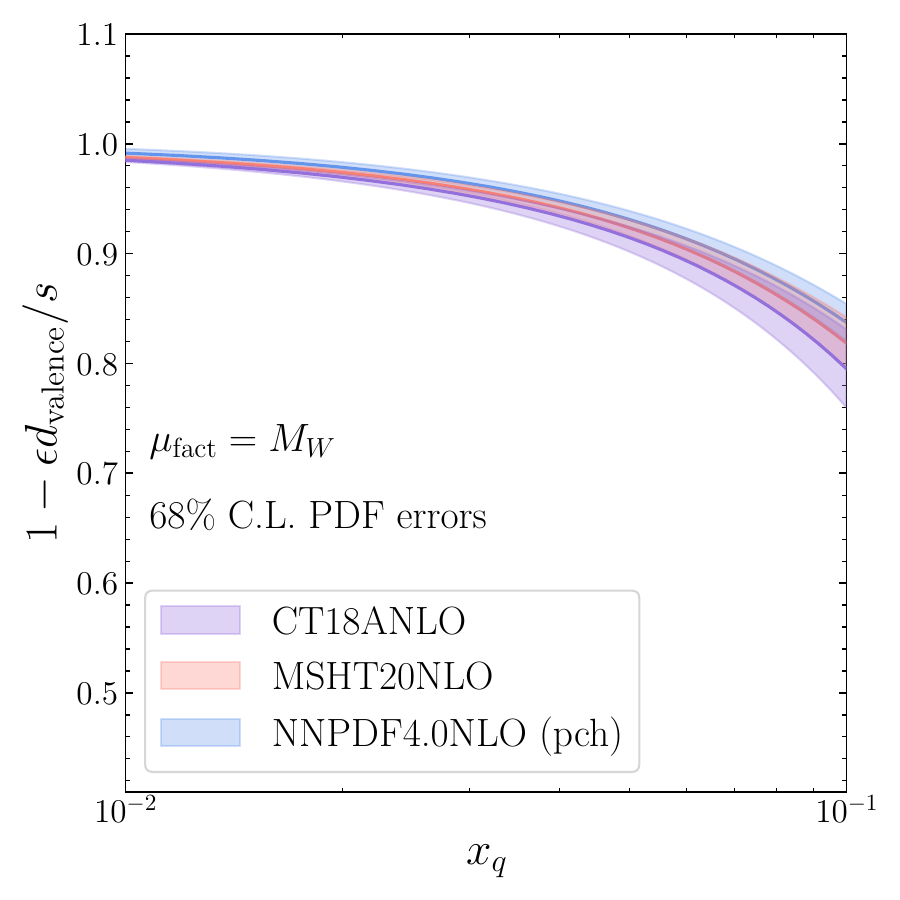}
        \caption{}
        \label{fig: Rcpm approximation sv0}
    \end{subfigure}
    \caption{\textbf{a)} 
    The combination of PDFs in Eq.~(\ref{eq:pcomb}) as a function of the momentum fraction $x_q$. \textbf{b)} Setting $s_\text{valence} = 0$ in in Eq.~(\ref{eq:pcomb}).
}
\end{figure}

We can use a simple approximation to argue that the above observations point to the non-zero $s_\text{valence}$ in MSHT20 and NNPDF4.0. The deviations in Figure~\ref{fig: Rcpm} grows as either $p_{T, D}$ or $|\eta_l|$ increases, so we expect this to reflect different mutual behaviour of PDFs as $x$ grows. As the gluon PDF grows fast at smaller values of $x$, we can further approximate the relevant events to stem from partonic interactions with a small-$x$ gluon and a large-$x$ quark. With this reasoning, the production ratio should reflect the following ratio of PDFs,
\begin{equation}
    \frac{|V_{cd}|^2\Bar{d} + |V_{cs}|^2\Bar{s}}{|V_{cd}|^2d + |V_{cs}|^2s} \approx 1 - \frac{\epsilon d_\text{valence} + s_\text{valence}}{s}, \quad \epsilon \equiv \frac{|V_{cd}|^2}{|V_{cs}|^2} \approx 0.05 \,,
    \label{eq:pcomb}
\end{equation}
where $V_{cd}$ refer to the elements of the CKM matrix. This quantity is plotted in Figure~\ref{fig: Rcpm approximation} as a function of the momentum fraction of the quark $x_q$. The curves are seen to loosely follow the pattern of theory predictions in Figure~\ref{fig: Rcpm}. In Figure \ref{fig: Rcpm approximation sv0} we have set $s_\text{valence} = 0$, leading to a good agreement between the three PDF fits. This suggests that it is indeed the non-zero $s_\text{valence}$ that causes the differences both in Figure~\ref{fig: Rcpm approximation} and Figure~\ref{fig: Rcpm}, and that decreasing $s_\text{valence}$ in MSHT20 and NNPDF4.0 at the relevant values of $x$ would bring the corresponding $R_c^\pm$ values closer to the data. This conclusion is further supported \cite{Alanko:2026hoc} by reweighting the PDFs with the data in Figure~\ref{fig: Rcpm}.
\section{Summary}
We have summarized a NLO-level GM-VFNS calculation of simultaneous production of a W boson and a charmed hadron. In particular, we have pointed out that the ratios between oppositely charged W bosons can provide a sensitive probe of the strangeness asymmetry near $x \sim 10^{-2}$ and that PDF fits with small asymmetry are favored by the recent ATLAS data. 

\acknowledgments
We acknowledge the financial support from the Research Council of Finland project 361179 and the Center of Excellence in Quark Matter of the Research Council of Finland, project 364194. The Finnish IT Center for Science, project jyy2580, is acknowledge for computing resources.

\bibliographystyle{JHEP}
\bibliography{references}

\end{document}